\documentclass[twocolumn,amsmath,amssymb,prl,showpacs,floatfix]{revtex4-1}
\usepackage{graphicx}
\usepackage[usenames, dvipsnames]{color}

\begin{document}

\title{Effects of classical stochastic webs on the quantum dynamics of cold atomic gases in a moving optical lattice}
\author{N. Welch$^{1}$, M.T. Greenaway$^2$ and T.M. Fromhold$^{1}$}

\address{$^1$Midlands Ultracold Atom Research Center, School of Physics \& Astronomy, University of Nottingham, Nottingham NG7 2RD, UK \\
$^2$Department of Physics, Loughborough University, Loughborough LE11 3TU, UK }

\begin{abstract}
We introduce and investigate a system that uses temporal resonance-induced phase space pathways to create strong coupling between an atomic Bose-Einstein condensate and a traveling optical lattice potential. We show that these pathways thread both the classical and quantum phase space of the atom cloud, even when the optical lattice potential is arbitrarily weak. The topology of the pathways, which form web-like patterns, can by controled by changing the amplitude and period of the optical lattice. In turn, this control can be used to increase and limit the BEC's center-of-mass kinetic energy to pre-specified values. Surprisingly, the strength of the atom-lattice interaction and resulting BEC heating of the center-of-mass motion is enhanced by the repulsive inter-atomic interactions.   
\end{abstract}


\maketitle{}

There is great interest in the interaction of harmonically-trapped atomic Bose-Einstein Condensates (BECs) with optical lattice (OL) potentials \cite{Mur2012, Cam2010, Kui2011, She2012, Zha2008}. Such systems have been used to realise and study a wide range of BEC dynamics including soliton formation and evolution \cite{Tsi2017}, the effect of nonlinear interactions on quantum tunneling \cite{Nes2014} and BEC transport through energy bands \cite{Oza2017}. Conversely, studies of the interactions of ultracold atoms with real crystal lattices, for example semiconductor surfaces, have highlighted the strong Casimir-Polder attraction of atoms to a surface that is approximately 1 $\mu$m away and shown how that attraction can facilitate interaction between atomic gases and condensed matter \cite{Ket2004,She2005,Fro2011}. For example, the Casimir-Polder attraction has been used to couple BECs in a harmonic trap to a vibrating SiN cantilever whose mean position is of order 1 $\mu$m away from the trap center \cite{Tre2010}. Dynamical perturbations have also been applied to BECs by using oscillating OLs \cite{Moh2012, Zha2010, Tin2010} and by exploiting Feshbach resonances to create a periodic driving term \cite{Ton2012}. All of these studies have shown that resonant driving can excite discrete modes within the BECs, which often induces center-of-mass motion of the atom cloud, thus heating it and causing atoms to be lost from harmonic traps.  

Perturbing a simple harmonic oscillator by a plane wave whose frequency is commensurate with the oscillator frequency has been shown rapidly to excite the oscillator \cite{chernikov1987minimal,zaslavski1992weak,shlesinger1993strange,fromhold2001effects,fromhold2004chaotic,fowler2007magnetic,balanov2008bifurcations,greenaway2009controlling,selskii2011effect,alexeeva2012controlling}. This resonant heating manifests itself experimentally in, for example, enhanced electron transport in semiconductor superlattices \cite{fromhold2001effects,fromhold2004chaotic,fowler2007magnetic,balanov2008bifurcations,greenaway2009controlling,selskii2011effect,alexeeva2012controlling} and heating in Tokamak fusion reactors \cite{Ber1977,chernikov1987minimal,zaslavski1992weak,shlesinger1993strange}. The excitation of the oscillator is due to the resonant creation of intricate phase space structures known as ``stochastic webs'' \cite{chernikov1987minimal,zaslavski1992weak,shlesinger1993strange,fromhold2001effects,fromhold2004chaotic,fowler2007magnetic,balanov2008bifurcations,greenaway2009controlling,Afr2011,selskii2011effect,alexeeva2012controlling}, which enable the oscillator to diffuse through the web filaments away from the web center, thereby gaining energy and becoming delocalised in real space. 

In this paper, we investigate the dynamics of a BEC with no, or repulsive, inter-atomic interactions, which is confined by a three-dimensional harmonic trap and driven by a traveling OL potential. We show that the stochastic web that forms resonantly in the single-atom classical phase space has a pronounced effect on the evolution of the BEC's density profile. In particular, the quantum-mechanical Wigner functions calculated for the condensate wavefunction diffuses through regions of phase space that correspond to the classical stochastic web filaments. This causes the Wigner functions to expand outwards away from the origin of the phase space, through channels aligned with radial filaments in the classical stochastic web, thereby increasing the BEC's center-of-mass motion. By adjusting the shape of the OL potential, and hence tailoring the underlying classical phase space structure, we can control \emph{a priori} the limit to which the BEC heats and derive an analytical model for the energy gained from the OL as a function of its wavelength. Although the heating rate increases with the strength of the OL potential, resonant diffusion of the BEC's Wigner functions along stochastic web filaments persists for arbitrarily small OL perturbations.

\section*{Semiclassical Modeling of the Perturbed BEC}

In our investigation we simulate the spatio-temporal evolution of a BEC using the Gross-Pitaeveskii equation (GPE)

\begin{equation}
i\hbar \frac{d\psi(\mathbf{x},t)}{dt} =\left [-\frac{\hbar^2}{2m}\nabla^2 + V(\mathbf{x},t) + g|\psi(\mathbf{x},t)|^2\right]\psi(\mathbf{x},t),
\end{equation} 

in which inter-atomic collisions are taken into account via the non-linear term, proportional to the local atom density, in the effective potential energy \cite{Fra2010, Mir2012, Adh2010, Shu2010}. Within the GPE, the distribution of atoms of mass, $m$, is described via a mean-field wavefunction, $\psi(\mathbf{x},t)$, and for the interacting cloud, the strength of the mean-field inter-atomic interaction potential is given by $g = g_0 = 4\pi\hbar^2a_{s}/m$, where $a_s$ is the s-wave scattering length. For non-interacting clouds, we will set $g = 0$. 

For BECs with repulsive inter-atomic interactions, $g_0$ is positive, meaning that the potential energy of the BEC increases with increasing atomic density. The total external potential energy in the GPE above, $V(\mathbf{x},t)$, comprises the harmonic trapping potential, $V_{trap}(\mathbf{x}) = m(\omega_x^2x^2 + \omega_y^2y^2 +\omega_z^2z^2)/2$, and the time-dependent OL perturbation that drives the atom cloud. The harmonic lengths along $x$, $y$, $z$ are defined by $l_{x,y,z}=(\hbar/m\omega_{x,y,z})^{1/2}$. We consider an OL traveling along the $z$ direction and described by the plane wave potential $V_{pert}(z,t) = V_o\cos(k_cz - \omega_ct)$, where  $V_o, k_c$ and $\omega_c$ are the wave amplitude, wavevector and angular frequency, respectively. We define the ratio $R = \omega_c/\omega_z$, which is a positive integer for temporally resonant driving. The OL wavevector, $k_c$ will, throughout, be scaled by the inverse of the harmonic trapping length, $l_z$.

To solve the GPE, we apply a projector to the mean-field wavefunction in order to create the Projected Gross-Pitaevskii equation (PGPE) \cite{Gar2008}. Physically, this projection limits the mean-field wavefunction to expansion over the cooler, highly-occupied, energy eigenfunctions of the 3D harmonic trap: see Ref. \cite{Gar2008} for details.

The PGPE provides a qualitatively and quantitatively accurate description of the dynamics of atomic gases that are sufficiently cold for their incoherent fraction to be neglected \cite{Bla2009}. In this regime, the PGPE has been shown to produce quantitatively accurate results when compared with experiments on atomic gases in, and far from, equilibrium \cite{New1} and when compared to more complex theoretical formalisms \cite{New2}.

However, there has also been much discussion of when and how these projected methods break down, for example in the case of quasi-condensate systems where the non-coherent part of the atomic cloud plays a much larger role \cite{New3}. In such cases, more powerful methods are required including the Stochastic Projected Gross-Pitaevskii Equation (SPGPE), Stochastic Gross-Pitaevskii Equation (SGPE) and Zaremba-Nikuni-Griffin (ZNG) formalism \cite{New4}. Even at very low temperatures, dynamical excitations can sometimes affect quantum-mechanical coherence across the cloud. However, it has been shown that even if coherence is partially lost, the results from simple GPE-like methods still agree with more complex methodologies for atoms in chaotic systems, see, for example \cite{New5}.

Here, we only consider low temperature regimes and atom cloud parameters chosen to ensure that decoherence and quantum-mechanical fluctuations \cite{Nor2005}, which only add 1/2 an atom to the $\sim$ 300 atoms that typically occupy each energy level, have a negligible qualitative and quantitative effect on the behavior of the driven BECs. Additionally, we limit the projector to ensure that for typical OL parameters the PGPE meets the validity requirement that high-energy states near the cut-off are all highly occupied \cite{Gar2008}.

We consider a BEC comprising $^{87}$Rb atoms with $m = 1.455\times10^{-25} $ kg and $a_s =  5.4\times10^{-9}$ m. 
The trapping frequencies are $\omega_x = 2\pi\times120$ Hz and $\omega_{y,z} = 2\pi\times30$ Hz \cite{Gar2008}, creating a 
``pancake"-shaped BEC with a width-to-height aspect ratio of 4, as shown in Fig. 1. Using these parameters, the harmonic length $l_z = 1.94\mu$m. The number of atoms, $N$, in the system is an important parameter as the nonlinear self-interaction potential scales with the atom density. For very large $N$, the repulsive inter-atomic forces strongly affect both the internal and center-of-mass dynamics of the atom cloud. However, We find that taking an experimentally-accessible value of $N = 10^4$ \cite{Gar2008} leads to an evolution that is influenced both by inter-atomic collisions and by the single-particle resonant phenomena, in particular the formation of extended stochastic web patterns in phase space, whose effect on BEC dynamics is the focus of this paper.

\begin{figure}
\centering
 \includegraphics*[width=0.8\linewidth]{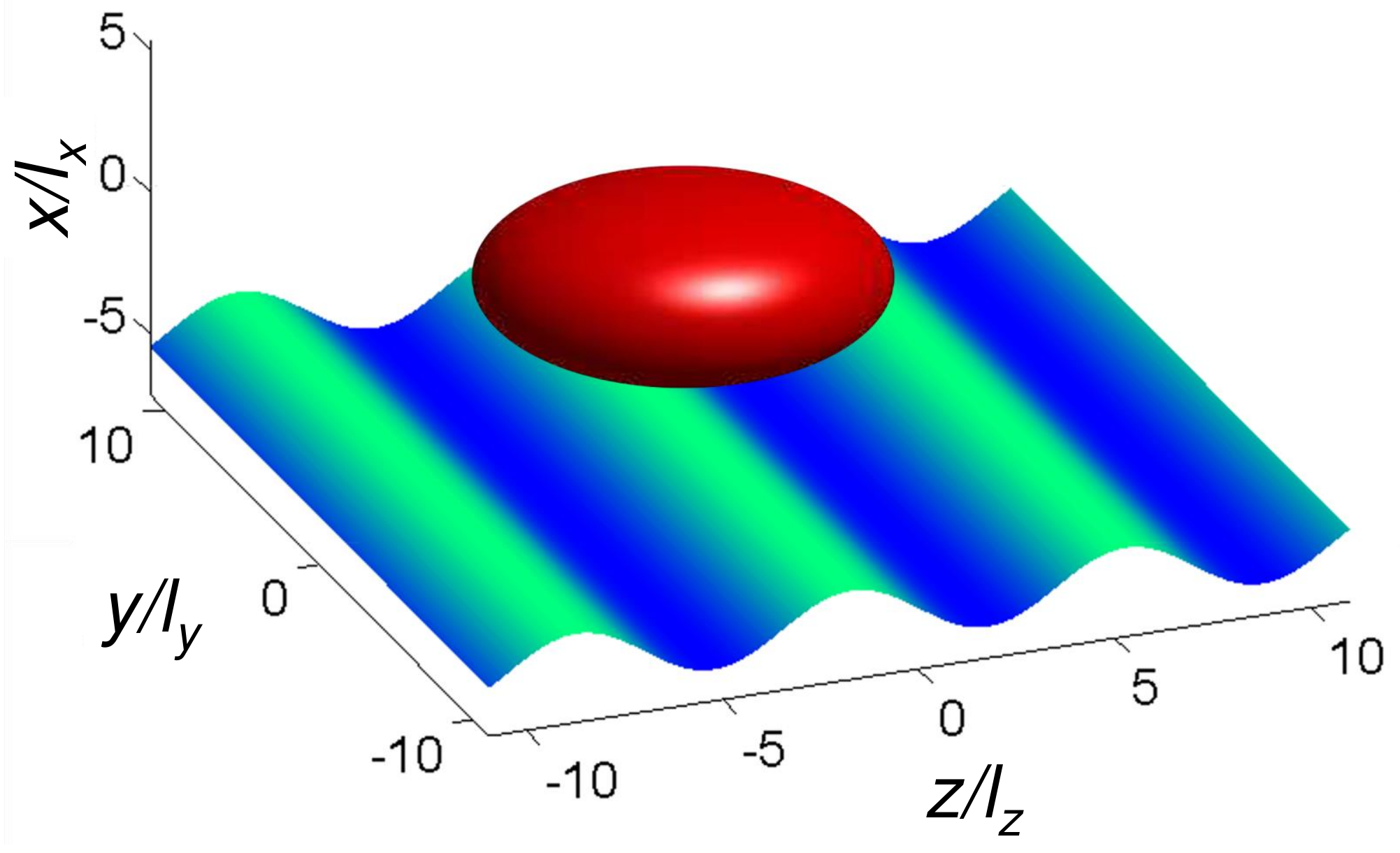}

   \caption{Iso-density surface (red) of a BEC that is confined in a 3D harmonic potential and subject to a plane wave potential (high/low potential energy shown green/blue) traveling along the $z$ direction, corresponding to the lowest frequency of the harmonic trap.} 
 \label{Fig.1}
\end{figure}

The initial state used in our quantum simulations is the groundstate of the harmonic trap, normalised so that $N= 10^4$, which we determine 
by evolving the PGPE using the imaginary time substitution $t\rightarrow-it$ so that excited-state contributions to the density profile decay exponentially leaving only the ground state. We also consider non-interacting atomic clouds ($g = 0$) in which every atom initially occupies the single-particle groundstate of the harmonic trap.

\begin{figure}
\centering
 \includegraphics*[width=1\linewidth]{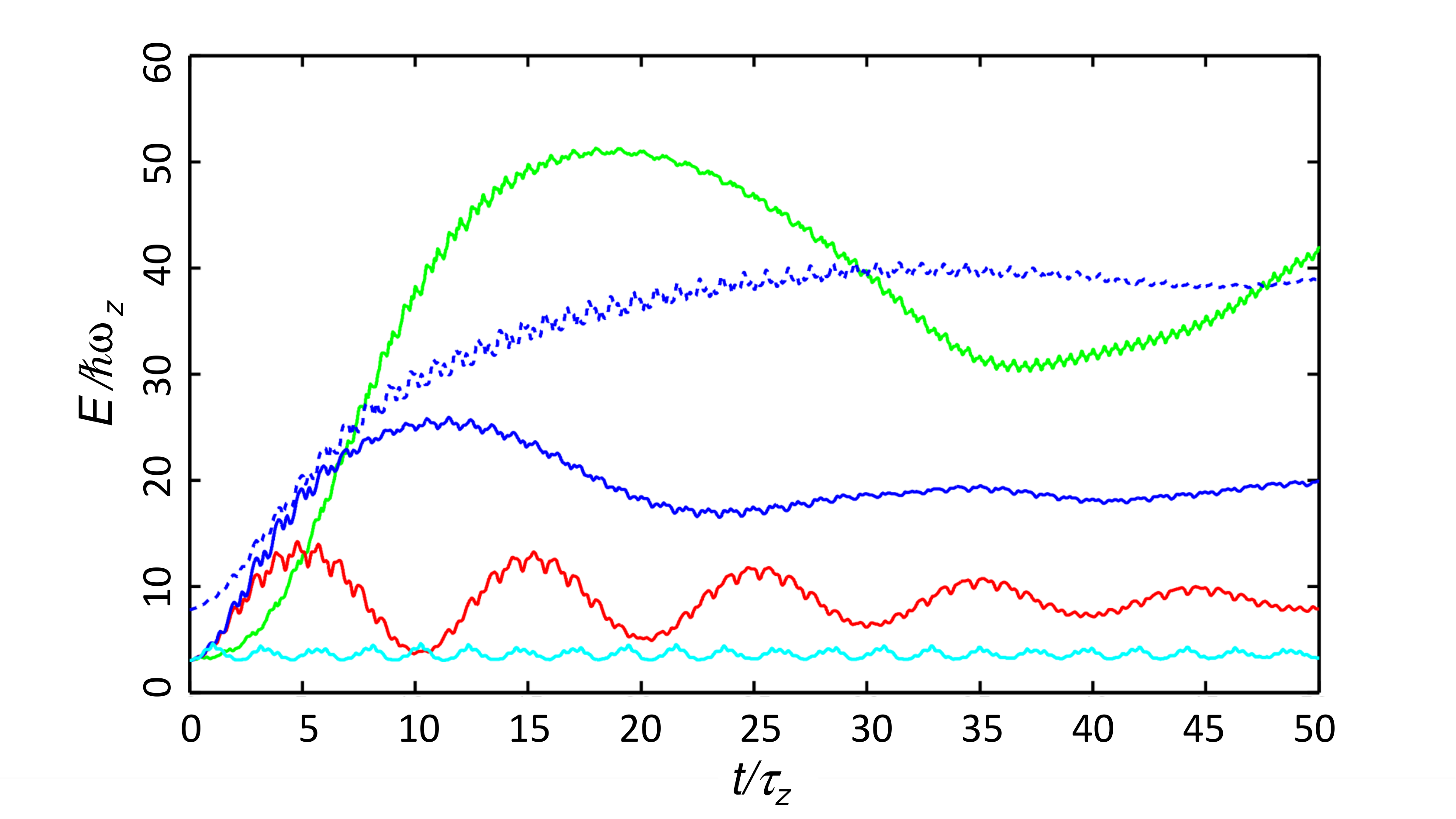}
   \caption{Comparison between the energy per atom, $E(t)$, in the BEC calculated versus time, $t$, shown in units of $\tau_z$, for resonant and non-resonant plane wave driving and for different strengths of inter-atomic interaction: Dark blue solid and dashed curves shows $R = 1$ resonant heating for non-interacting $(g=0)$ and interacting atoms respectively; red and light blue curves show non-resonant heating for $R = 0.95$ and $R = \sqrt{2}$ respectively ($g=0$ in both cases); green curve shows $R = 2$ heating for $g=0$ and $V_o = 1.3\hbar\omega_z$. $V_o = 0.3\hbar\omega_z$ for all other curves. In all cases, $N = 10^4$ and $k_c = 0.5/l_x$.}
 \label{Fig.2}
\end{figure}

Even though the PGPE only describes a micro-canonical ensemble, it can still be used to determine the temperature of the atom cloud in equilibrium \cite{Gar2008, Rug1997}. However, since a BEC driven by a traveling OL is heated dynamically and evolves far from equilibrium, its temperature is not well defined. Consequently, we use the energy per atom, $E(t)$,
\begin{equation}
E(t) =\frac{1}{N} \int^{\infty}_{-\infty}\psi^\dag(\mathbf{x},t)i\hbar \frac{d\psi(\mathbf{x},t)}{dt}\mathrm{d}\mathbf{x},
\end{equation}
as the measure of heating and to quantify the instantaneous energy of the atom cloud.  

\begin{figure*}
 \includegraphics*[width=\textwidth]{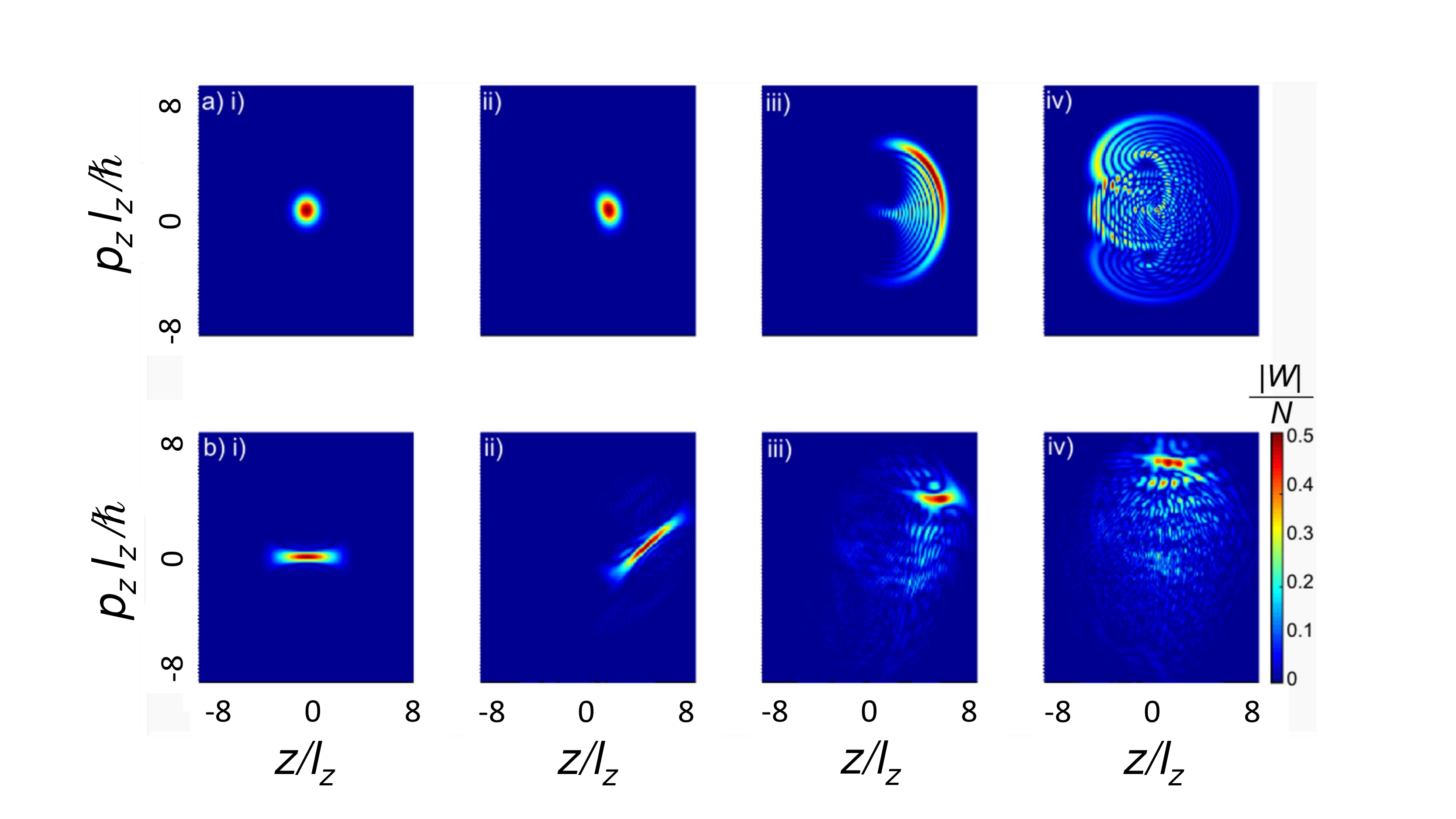}

   \caption{(a, i-iv) Position-momentum $(z,p_z$) phase space Wigner functions (normalised moduli, $\lvert W \rvert/N$, plotted) showing the time evolution calculated for a non-interacting atom cloud with resonant $R=1$ plane wave driving. Panels show: (i) $t = 0$, cloud at rest; (ii)$t = \tau_z$, cloud has been excited to a larger phase space radius, $\rho = \sqrt{z^2 + p_z^2}$ along the $\langle p_z \rangle = 0$ pathway; (iii) $t = 7\tau_z$, cloud reaches, and scatters from, a ring-shaped dynamical barrier; (iv) $t = 17\tau_z$, cloud begins to contract along the $\langle p_z \rangle = 0$ pathway, reducing its mean $\rho$ value and energy. (b, i-iv) Phase space evolution calculated for an interacting atom cloud ($g=g_0$) at times: (i) $t = 0$, cloud at rest but repulsive interactions produce a larger initial phase space spread than for $g=0$; (ii)$t = 6\tau_z$, cloud is excited as in Panel (a,ii) except that the interactions have distorted the phase space distribution; (iii)$t = 17\tau_z$, the cloud reaches the ring-shaped dynamical barrier and scatters from it, becoming fragmented; (iv) $t = 25\tau_z$, cloud continues to evolve in phase space with no reduction in its energy. Color bar shows normalised values of the Wigner function modulus.} 
 \label{Fig.3}
\end{figure*}

Figure 2 shows $E(t)$ curves calculated versus time, $t$, in units of $\tau_z=2\pi/\omega_z$, for various optical lattice parameters to highlight the enhanced heating that occurs when the harmonic trap is driven resonantly. The two dark blue curves show $E(t)$ for non-interacting ($g = 0$, solid dark blue curve) and interacting ($g=g_0$, dashed dark blue curve) atom clouds when $R = 1$. Both curves have approximately the same form for $t<10$, although when $g>0$ the atom cloud has a larger initial energy due to the repulsive inter-atomic interactions. For $t \gtrsim 10$, $E(t)$ attains a maximum value of $E\approx25\hbar\omega_z$ when $g=0$, but thereafter decreases. By contrast, for $g=g_0$, $E(t)$ reaches a maximum and then remains approximately constant. The red and light blue curves in Fig. 2 show $E(t)$ for non-resonant heating, with $R = 0.95$ and $\sqrt{2}$ respectively, when $g=0$. Both curves show that $E(t)$ increases far slower off resonance, even when $R=0.95$, than for the $R=1$ resonant case. The solid green curve in Fig. 2 shows $E(t)$ calculated for $R = 2$, and shows that the energy of the atom cloud increases to a higher maximum, than when $R=1$. However, in order to reach the higher maximum in approximately the same time, the optical lattice depth is increased by more than a factor of 4 when $R=2$ compared to $R=1$.

To understand the evolution of $E(t)$, we now consider the quantum phase space evolution of the atom cloud by calculating the Wigner quasi-probability distribution function using
\begin{equation}
W(\mathbf{x},\mathbf{p}) = \frac{1}{\pi\hbar} \int^\infty_{-\infty}{\psi^\dag(\mathbf{x}+\mathbf{\Lambda})\psi(\mathbf{x}-\mathbf{\Lambda})
e^{2i\mathbf{p}.\mathbf{\Lambda}/\hbar}\mathrm{d}\mathbf{\Lambda}},
\end{equation}
where $\mathbf{p} = p_x\mathbf{\hat{x}} +  p_y\mathbf{\hat{y}}  +  p_z\mathbf{\hat{z}}$  and $\mathbf{\Lambda} = \lambda_x\mathbf{\hat{x}} +  \lambda_y\mathbf{\hat{y}}  +  \lambda_z\mathbf{\hat{z}}$. To reduce this six-dimensional distribution to a two-dimensional function of the variables $z$ and $p_z$ that characterise motion along the OL direction, we integrate over the other four dimensions. 
Due to its form, $W(z,p_z)$ is real, therefore the accuracy of the integration can be quantified by the ratio of the resulting imaginary and real parts. In our simulations we ensure that Im$(W)/$Re$(W)<10^{-10}$.  

Figure 3(a,i-iv) shows the Wigner function evolution calculated for a non-interacting cloud when $R = 1$, i.e. corresponding to the solid blue curve in Fig. 2. Initially, the probability distribution is centered on $(z,p_z)=(0,0)$, corresponding to the harmonic trap ground state (Fig. 3a i). As $t$ increases, the OL excites the BEC into oscillatory center-of-mass motion. After the OL has acted for one trap period, $\tau_z$, the atom cloud has moved along the $z$ axis, increasing its potential energy in the harmonic trap. This can be seen from the Wigner function in Fig. 3(a,ii), whose horizontal displacement and symmetry reveal that the time averaged $z$ co-ordinate $\langle z\rangle > 0$ and that the atom cloud has almost zero average momentum ($\langle p_z\rangle \approx 0$). In addition, the internal form of Wigner function is similar to that at $t=0$, which means that the BEC's spatial shape is preserved. As $t$ further increases, the Wigner function continues to diffuse along the $z$ axis and $p_z=0$ pathway, and the BEC moves further away from the harmonic trap center, gaining potential energy. Eventually, the BEC's Wigner function reaches a region of phase space, determined by the wavevector of the OL as explained below, where it no longer moves to higher $z$ values but, instead, broadens and moves away from $p_z=0$ around a ring-like dynamical barrier of radius $\rho = \sqrt{z^2 + p_z^2}$ in phase space [Fig. 3(a,iii)]. Next, the Wigner function profile remains ring-like but begins to move back towards $(z,p_z)=(0,0)$, decreasing its phase-space radius and, hence, causing $E(t)$ to decrease as shown for $t>10\tau_z$ by the dark blue solid curve in Fig. 2. 

Figures 3(b,i-iv) show the Wigner function evolution calculated for the BEC with repulsive inter-atomic interactions. Initially [Fig. 3(b,i)], the atom cloud has a larger spread in position than when $g=0$ [Fig. 3(a,i)]. As $t$ increases to $6\tau_z$ [Fig. 3(b,ii)], the OL driving potential causes the BEC's Wigner function to travel along the same $\langle p_z \rangle = 0$ pathway as for $g=0$ [Fig. 3(a,ii)]. However, for the BEC, the repulsive inter-atomic interactions strongly affect the form of the Wigner function and cause the atom density profile to spread and fragment in both phase and real space. Figure 3(b,iii) captures, at $t=17 \tau_z$, the Wigner function moving out around the same circular phase space barrier as for $g=0$ [Fig. 3(a,iii)]. However, in contrast to the almost symmetric spreading seen for $g=0$ [Figs. 3(a,iii-iv)], the BEC's Wigner function becomes highly fragmented as it reflects from the circular dynamical barrier and expands rapidly in both $z$ and $p_z$ [Fig. 3(b,iv)]. All of the Wigner functions shown in Fig. 3 for $t>0$, reveal asymmetry between $+p_z$ and $-p_z$ because propagation of the OL traveling wave along the positive $z$ direction makes the atom cloud drift in this direction. This asymmetry diminishes as the OL amplitude, $V_o$, decreases. 

The repulsive $g|\psi(\mathbf{x},t)|^2$ potential breaks the position-momentum symmetry that is embedded in the equation of motion for the non-interacting atom cloud. The effect of this symmetry breaking on the evolution of the atom density profile is particularly pronounced when the Wigner function begins to move around the circular dynamical barrier in phase space [e.g. Fig. 3(b,iii)] because the atom cloud decelerates along $z$ thus increasing the peak atom density. The symmetry breaking prevents the interacting cloud from contracting back towards $(z,p_z)=(0,0)$ along the $\langle p_z \rangle = 0$ pathway following reflection from the dynamical barrier when $t \sim 15\tau_z$. This explains why, for the BEC, $E(t)$ continues to increase for $t\gtrsim 20\tau_z$ (dark blue dashed curve in Fig. 2) in contrast to the energy loss seen when $t\gtrsim 10\tau_z$ for the non-interacting cloud (dark blue solid curve Fig. 2).



To gain insight into the form and evolution of the Wigner functions, we now consider the corresponding classical equations of motion. A single harmonically-confined atom with no inter-atomic collisions performs simple harmonic motion along the $x$ and $y$ directions, which separates from that along $z$. Motion along the $z$ direction corresponds to a harmonic oscillator driven by the traveling OL potential and is described by the following equation:
\begin{equation}
\ddot{z} = -\omega_{z}^{2}z + \frac{V_ok_c}{m}\sin(k_cz - \omega_ct),
\end{equation}
where all of the parameters take the numerical values described above in order to facilitate direct comparison between classical and quantum evolution of the atom cloud. To determine the classical dynamics, we solved Eq. (4) using the 4th order Runge-Kutta algorithm. The initial conditions were taken to be a set of pseudo-randomly created positions and momenta with Gaussian distributions, corresponding to those of the single-particle quantum-mechanical groundstate of the harmonic potential along $z$ and with the same associated energy, $\hbar\omega_z/2$. 

Figure 4(a) shows a stroboscopic Poincar\'{e} section constructed by plotting the phase space variables $(z,p_z)$ obtained from Eq. (4) at equally-spaced times $t=l\tau_z, l=0,1,2,...$, for $R = 1$. The form of the Poincar\'{e} section differs markedly from the elliptical islands of stability found in the absence of the plane wave driving potential. In particular, there are two distinct islands with an almost linear separatrix (blue dashed line) and enclosed by a circular boundary (red dashed curve). Around these dashed regions of phase space, the classical motion is chaotic and the width of the chaotic regions increases with the amplitude of the plane wave perturbation in Eq. (4). 

\begin{figure}
\centering
 \includegraphics*[width=1.1\linewidth]{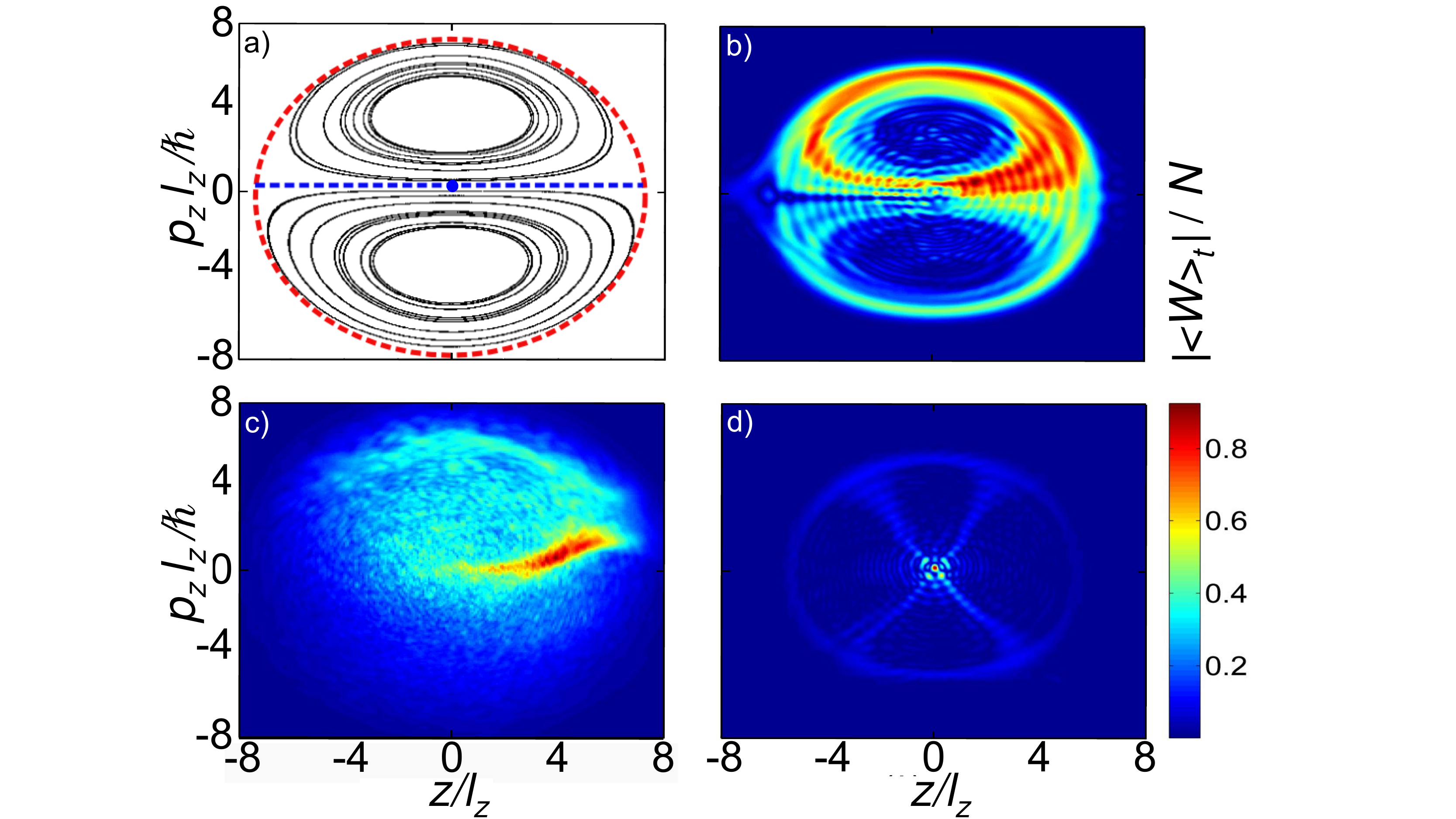}

 \caption{(a) Classical Poincar\'{e} section (dots) calculated for an $^{87}$Rb atom driven by a plane wave. Blue and red dashed curves show, respectively, the location of the radial and first ring-shaped filaments of the stochastic web that encloses the islands of stability shown black. (b)-(d) Wigner functions (normalised moduli plotted: color bar lower right) periodically time-averaged over $40\tau_z$ for $R=1$ with $g=0$ (b) and $g=g_0$ (c) and for $R = 2$ with $g=0$ (d). For all panels, $V_o = 0.3\hbar \omega_z$. For (a-c) $k_c = 0.5/l_z$ and for (d) $k_c = 0.75/l_z$.} 

 \label{Fig.4}
\end{figure}

It is well known that the phase space of a harmonic oscillator driven by a plane wave whose angular frequency is an integer multiple of the unperturbed harmonic oscillator frequency is threaded by a continuous stochastic web within which the motion is unstable \cite{chernikov1987minimal,zaslavski1992weak,shlesinger1993strange,fromhold2001effects,fromhold2004chaotic,fowler2007magnetic,balanov2008bifurcations,greenaway2009controlling,Afr2011,selskii2011effect,alexeeva2012controlling}. Each stochastic web is characterised by an infinite number of circular filaments connected by 2$R$ radial filaments. Fig. 4(a) shows the position of the 2 radial filaments (blue dashed lines) that extend outwards along $p_z=0$ and the innermost ring-shaped filament (dashed red curve). Stochastic webs have analytically-defined structures that can be determined by expanding the Hamiltonian of the driven harmonic oscillator into Bessel functions and finding the stationary points of the system \cite{Afr2011}. The concentric ring-shaped filaments have radii given by $\rho^N_R = A^N_R/k_c$, where $A^N_R$ is the $N^{th}$ root of the $R^{th}$ Bessel function, $J_R$, i.e. where $J_R(A^N_R) = 0$. The dashed red ring shown in Fig. 4 corresponds to $N = 1$ and $R = 1$. 

Stochastic webs are important in many branches of physics including, plasma dynamics, Tokamak fusion, quasi crystals, condensed matter quantum devices and analogous optical and atomic systems \cite{chernikov1987minimal,zaslavski1992weak,shlesinger1993strange,fromhold2001effects,fromhold2004chaotic,fowler2007magnetic,balanov2008bifurcations,greenaway2009controlling,Afr2011,selskii2011effect,alexeeva2012controlling}. One reason for this interest is that the radial filaments provide chaotic pathways along which the oscillator can diffuse, thus attaining higher $\rho$ values and gaining energy from the plane wave. The existence of stochastic webs in the classical phase space of a harmonically-trapped atom resonantly driven by a traveling OL explains key features of the heating rates determined from quantum-mechanical models for the evolution of the atom cloud (Fig. 2). 

To explain this classical-quantum correspondence, we now consider time averages of a series of Wigner functions calculated every trap period, $\tau_z$, over the interval $0\leq t\leq 40\tau_z$. This time averaging makes the phase-space evolution of the atom density distribution clearer and facilitates comparison with stroboscopic Poincar\'{e} sections, like that shown in Fig. 4(a). Figure 4 also shows time-averaged Wigner functions calculated by solving the PGPE for $R=1$ and (b) $g=0$, (c) $g=g_0$ and for (d) $R=2$ and $g=0$. 

Although, in principle, the web filaments extend outwards to infinity, their width becomes exponentially thinner with increasing phase space radius, $\rho = \sqrt{z^2 + p_z^2}$. Consequently, in a classical picture, an atom starting from rest at the web center moves out along the $z>0$ (blue) radial filament in Fig. 4(a) and then, rather than continuing to move radially away from the web center,  transfers to the inner ring-shaped filament (red), before eventually re-entering the radial filament and moving back towards the web center. Trajectories in the two islands of stability enclosed by the web filaments have a similar form characterised by gaining and losing kinetic energy as they move respectively away from, and towards, the web center. The time-averaged Wigner function calculated for the non-interacting atom cloud [Fig. 4(b)] concentrates around the radial and innermost ring filaments of the stochastic web, indicating that the quantum-mechanical evolution of the atom cloud is shaped by the underlying classical dynamics. When inter-atomic interactions are included [Fig. 4(c)], the time-averaged Wigner function peaks near the $z>0$ radial filament and is bounded by the innermost ring filament. However, within that circular boundary the probability distribution is more diffuse than for $g=0$ because the inter-atomic interactions create additional forces that enable the driven oscillator to spread further into the islands of stability that are enclosed by the stochastic web. In the absence of interactions, the islands of stability are inaccessible to a classical trajectory starting from rest at the web center and so the corresponding time-averaged Wigner function [Fig. 4(b)] penetrates less far into the islands of stable classical motion.

Localisation of the classical orbits within the inner-most ring-shaped filament in Fig. 4(a) explains why, in a quantum picture, the atom cloud backscatters entirely from the first phase space ring, with no part of the atom cloud being able to travel further outwards beyond this ring, which therefore acts as a dynamical barrier in phase space. Since the ring radii have the analytical form given above, the corresponding maximum attainable energy per particle in the atom cloud is

\begin{equation}
E^R_{ring} = \frac{m\omega_z^2[A^1_R]^2}{2k_c^2}.
\end{equation}

As $R$ increases, the radius of the inner ring also increases as it corresponds to the first root of successively higher-order Bessel functions, with $A^1_1<A^1_2<A^1_3$ etc. This makes more of the phase space accessible to the driven oscillator as $R$ increases, which explains why the maximum energy that the atom cloud attains in Fig. 2 is higher for $R=2$ (green curve) than for $R=1$ (dark blue solid curve). However, with increasing $R$, the ability to excite the entire atom cloud along the radial phase space pathways to reach the innermost circular web filament diminishes because the web radius, $\rho_{R}^{1}$, increases \cite{Afr2011}. This can be seen from the time-averaged Wigner function calculated for $R=2$ [Fig. 4(d)]. Although this Wigner distribution is clearly shaped by the innermost circular filament and by the four (i.e. $2R$) radial filaments of the corresponding classical stochastic web, the time averaged probability density remains concentrated towards the center of the web at $(z,p_z)=(0,0)$ as it takes far longer for the cloud to reach the ring-shaped web filament when $R=2$ than for $R=1$. Indeed, as shown in Fig. 2, when $R=2$ (green curve) the optical lattice depth has to be increased approximately four-fold in order for the atom cloud to attain its maximum energy at a similar time to the $R = 1$ case (dark blue solid curve). Consequently, the most efficient excitation and heating of the cloud occurs when $R = 1$ rather than for higher frequency ratios.

\begin{figure}
\centering
 \includegraphics*[width=1\linewidth]{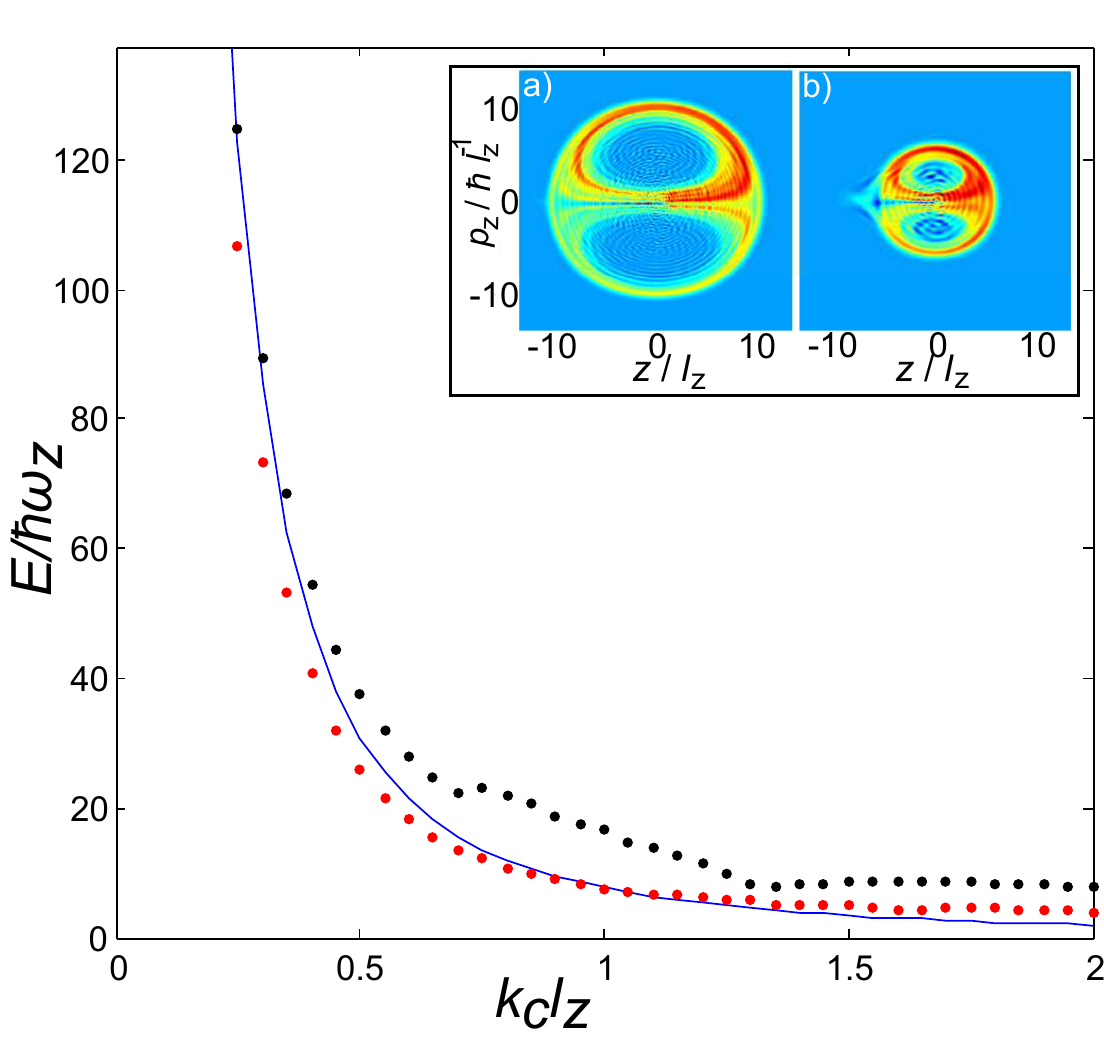}

   \caption{Maximum energy of atom clouds calculated versus wave vector, $k_c$ (shown multiplied by $l_z$), of the plane wave driving potential. Blue curve shows $E^{1}_{ring}$ values given by Eq. (5). Black (red) symbols are maximum energies calculated using the GPE with $g=g_0$ ($g=0$). Insets: Wigner functions (moduli plotted) calculated for $R=1$ and $k_c$ = (a) $0.4/l_z$, (b) $0.75/l_z$ peak around the classical stochastic webs in phase space.} 
 \label{Fig.5}
\end{figure}

The classically-derived Eq. (5) suggests that a traveling OL with $R=1$ can be used to heat a BEC to an energy, $E^1_{ring}$, that is selected \emph{a priori}, by simply varying $k_c$. To investigate the validity of Eq. (5) for predicting quantum dynamics, in Fig. 5 we compare $E^1_{ring}$ calculated versus $k_c$ from this equation (solid blue curve) with the maximum energy attained by interacting (black circles) and non-interacting (red circles) atom clouds whose quantum evolution is described by the PGPE. 

Although the quantum-mechanical results are generally very close to the classical prediction from Eq. (5), there are also some interesting differences. Primarily, the maximum energy attained by the interacting atom cloud (black circles) generally exceeds the classical prediction because the initial repulsive self interaction causes expansion of the atom cloud, thus increasing its kinetic energy. In addition, in contrast to the classical model, the maximum energy attained by the interacting atom cloud does not vary monotonically with $k_c$, revealing peaks around $k_c = 0.8/l_z$ and $1.6/l_z$. These peaks originate from spatial resonance between the width of the interacting atom cloud and the wavelength of the OL when $\lambda_c  = \frac{2\pi}{k_c} \sim 2z_{TF}$ and $\sim z_{TF}$ respectively, where $z_{TF}$ is the Thomas-Fermi radius, i.e. half the width of the atom cloud. When this spatial resonance condition is satisfied, the cloud acts more like a single atom, ensuring the entirety of the cloud is heated coherently, rather than fragmenting and converting its inter-atomic interaction energy into kinetic energy. When $k_c \gtrsim 1.25/l_z$, the maximum temperature reached by both the interacting and non-interacting atom clouds exceed the classical prediction. This is because, in this high $k_c$ regime, the radius of the inner stochastic web ring is smaller than the initial position and momentum spread of the two atom cloud groundstates. As a result, the corresponding Wigner functions can rapidly expand beyond the first web ring, causing the energy to increase above the maximum value predicted by Eq. (5).

The OL amplitude, $V_o$, also affects the heating of the atom cloud.  Although the stochastic web structure still forms resonantly in phase space for arbitrarily small $V_o$ values, the width of the web filaments, and hence the rate at which the atom cloud moves outwards along them in phase space, and how far it travels, both increase with increasing $V_o$. The color map in Fig. 6 shows the total energy of an interacting $^{87}$Rb BEC with $g=g_0$ calculated versus $t$ and $V_o$ for fixed $k_c = 0.5/l_z$. The dashed curve in the projected (lower) color map, which lies within the yellow band in the color scale, marks where the energy of the atom cloud reaches 90$\%$ of the maximum energy given by Eq. (5). As $V_o$ increases, the atom cloud reaches this energy threshold faster because, qualitatively, the stochastic web filaments are wider and hence support faster and further phase space diffusion. Quantitatively, analysis of the locus of the dashed curve shows that the time taken for the atom cloud to reach 90$\%$ of the maximum energy is $\tau_{heat} \approx 2.8\hbar/V_o$. The uneven nature of the heating shown in Fig. 6 is due partly to the self-interaction energy and partly to variation in the rate at which the atom cloud diffuses through phase space as its spatial and momentum distributions change \cite{Afr2011}. These effects limit the maximum heating rate, as seen in Fig. 6 for $V_o>3.5\hbar\omega_z$, where the time evolution of the BEC's energy depends only weakly on $V_o$. 

\begin{figure}
\centering
 \includegraphics*[width=1\linewidth]{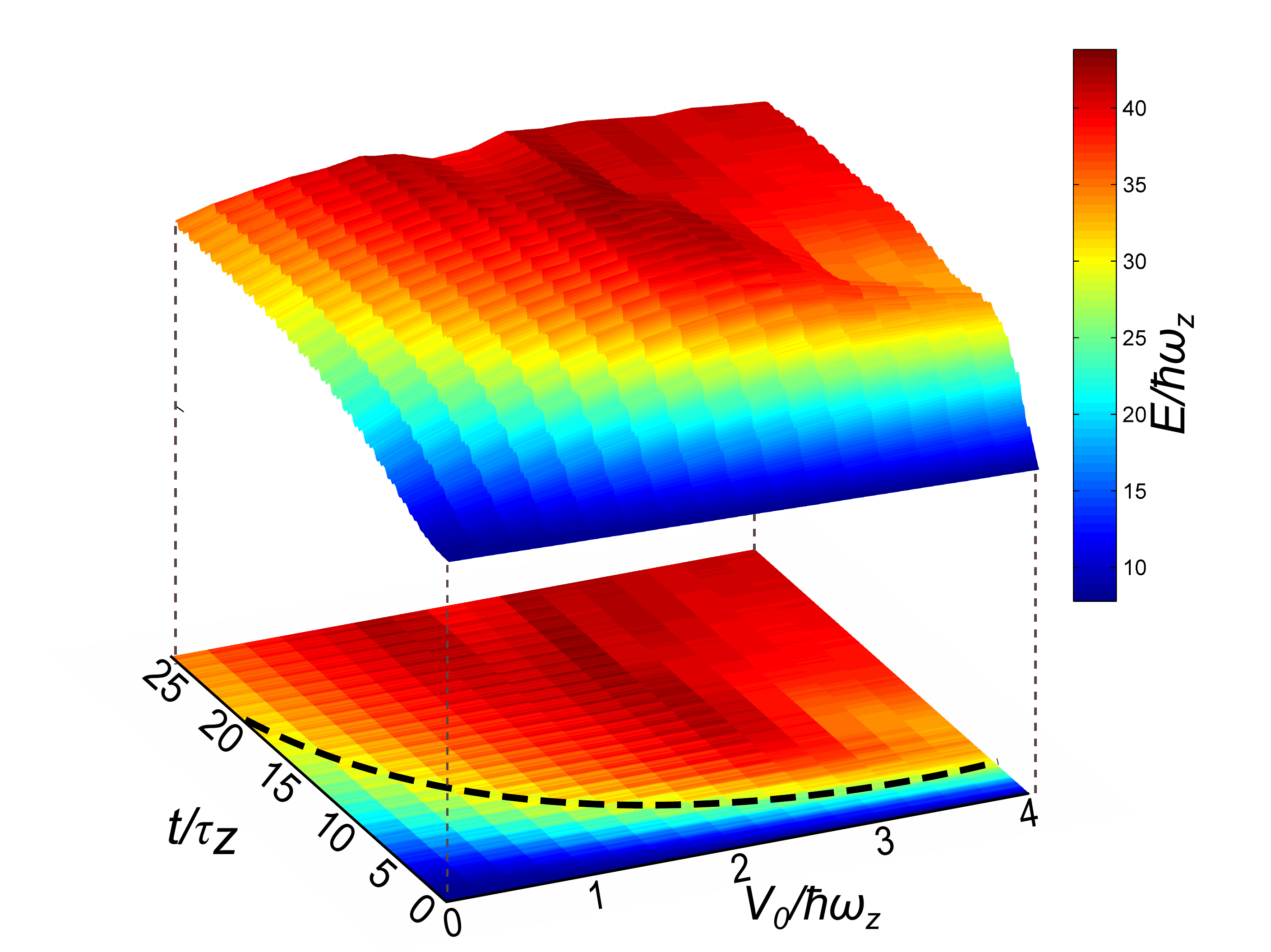}
   \caption{Color map (scale right) showing the total energy, $E$, of an $^{87}$Rb BEC calculated versus time (in units of $\tau_z$) and the amplitude, $V_o$ in units of $\hbar \omega_z$, of a plane wave driving potential with $k_c = 0.5 l_z$. The map is shown both as a surface plot (upper) and a 2D projection (lower). Along the dashed curve in the projection, the BEC's energy is 90\% of the energy limit given by Eq. (5).} 

\label{Fig.6}
\end{figure}

\begin{figure}
\centering
 \includegraphics*[width=1\linewidth]{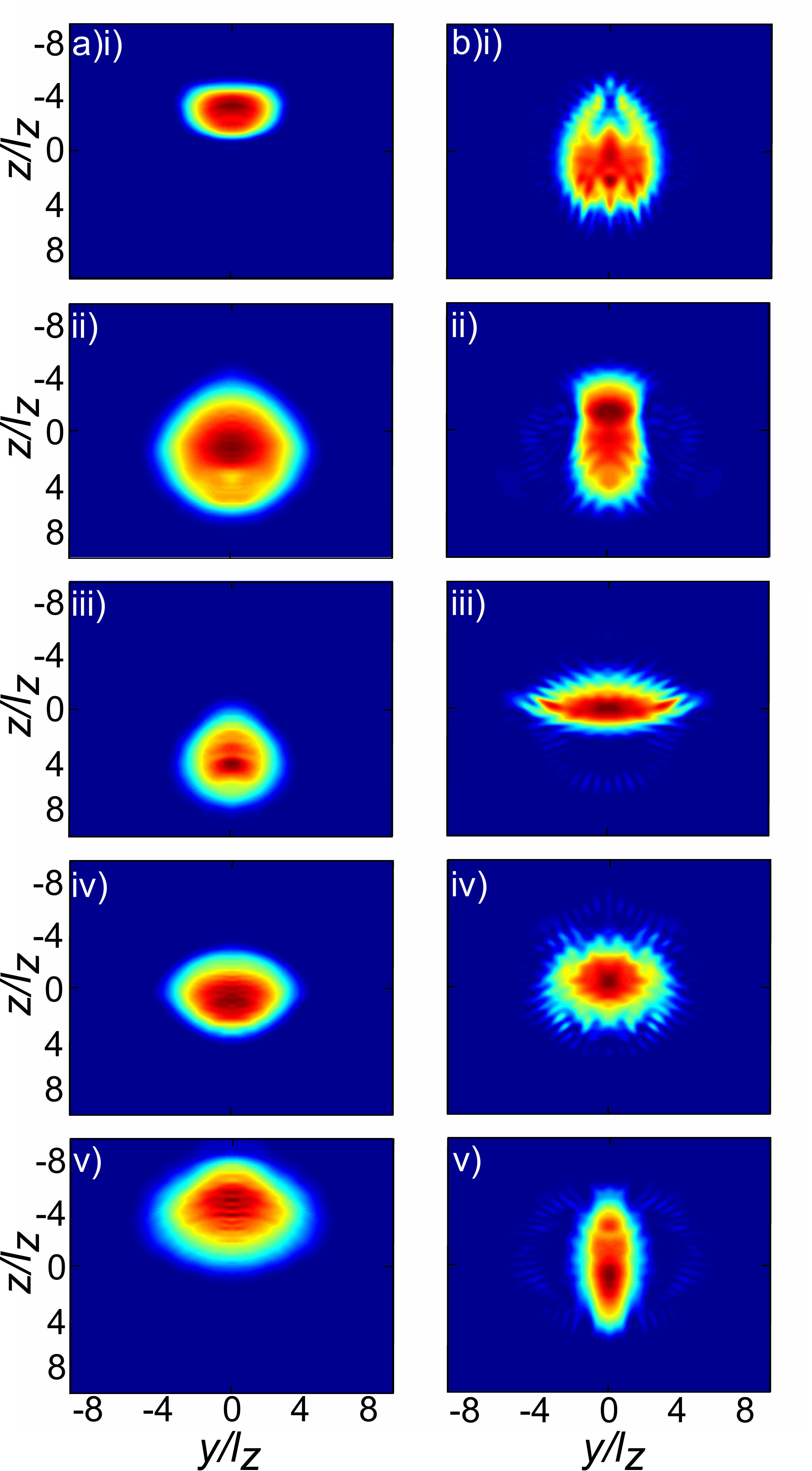}

   \caption{Atom density profiles calculated for an $^{87}$Rb BEC with $g=g_o$ shown in the $x = 0$ plane for $R = $ (a) 1, (b) 2 at times, $t =$ (i) $5.00\tau_z$, (ii) $5.25\tau_z$, (iii) $5.50\tau_z$, (iv) $5.75\tau_z$, (v) $6.00\tau_z$. Panels (a,i-v) show large-amplitude center-of-mass oscillation during which the BEC maintains a spatially-coherent form. In panels (b,i-v) there is no center-of-mass oscillation, but significant deformation and fragmentation of the atom cloud.}
 \label{Fig.7}
\end{figure}

The OL parameters can be controled to create a phase space stochastic web filaments that are wide enough to allow coherent center-of-mass motion in which the condensate gains energy with very little fragmentation. Such motion is shown in Fig. 7(a,i-v) (i.e. left-hand column) over a single harmonic trap period at times $t =$ (i) $5.00\tau_z$, (ii) $5.25\tau_z$, (iii) $5.50\tau_z$, (iv) $5.75\tau_z$, (v) $6.00\tau_z$. The atom cloud has a similar form at the start and end of the period and shows no internal fragmentation. By contrast, non-resonant heating, as shown in Fig. 7(b,i-v) (i.e. right-hand column), causes mainly internal deformation and the excitation of quadrupole oscillation modes within the atom cloud. In this case, the OL produces very little center-of-mass motion and, as a result, limited heating.

\section*{Conclusion}

Driving a harmonically-confined cold atom cloud with a traveling OL whose frequency is commensurate with that of the harmonic trap causes controlable heating of the cloud by imprinting a stochastic web in the classical phase space. The web filaments provide a network of transport channels through which the atoms can diffuse and gain energy in both classical and quantum descriptions of the atomic dynamics. The heating of the cloud can be selected \emph{a priori} by changing the wavelength of the OL, even for very small OL amplitudes. This result opens up many avenues for future work because controled excitation and heating of atomic gases typically requires strong coupling to a driving potential, for example via large deformations of the harmonic trapping potential. By contrast, we have shown that plane wave perturbations of arbitrarily small amplitude can exert a large, controlable effect on BECs. Delicate control potentials of this form could therefore be used for precise, non-destructive, quantum state preparation and coherent evolution. 
Our results also suggest new directions for experimental studies of BECs coupled to an oscillating cantilever or membrane \cite{Tre2010}. In particular, stochastic webs may created and used to control atom clouds by coupling the atoms, via the atom-surface Casimir-Polder attraction, to mechanical standing waves along the nearby solid state oscillator. Since resonant coupling has been shown to create stochastic webs in the atomic phase space \cite{Afr2011}, we expect it to cause rapid controlable increase in the kinetic energy corresponding to the BEC's center-of-mass motion and, conversely, sympathetic cooling of the solid-state oscillator. In principle, the oscillator could be cooled close to its quantum-mechanical groundstate by transferring energy to the atom cloud, thereby providing access to a range of macroscopic quantum phenomena in condensed matter \cite{Zan2012}. Moreover, such strong resonant coupling between surfaces and BECs, resulting from arbitrarily weak perturbations, could provide new mechanisms and protocols for coherent state transfer and readout.  

Acknowledgements: This work is supported by EPSRC UK.

\end{document}